\documentclass[epj]{svjour}
\usepackage{a4wide}
\usepackage{graphicx}
\usepackage{axodraw}
\usepackage{amssymb}

\usepackage{hyperref}

\newcommand{\be}{\begin{equation}}
\newcommand{\ee}{\end{equation}}
\newcommand{\ba}{\begin{eqnarray}}
\newcommand{\ea}{\end{eqnarray}}
\newcommand{\chitoPP}{\ensuremath{\chi_{c0},\chi_{c2}\to PP}}
\begin{document}

\onecolumn
\thispagestyle{empty}
\setcounter{page}{0}
\begin{flushright}
LU TP 11-37\\
September 2011
\end{flushright}
\vfill
\begin{center}
{\Large\bf Chiral Symmetry and Charmonium Decays to Two Pseudoscalars}
\vfill
{\bf Johan Bijnens and Ilaria Jemos}\\[0.3cm]
{Department of Astronomy and Theoretical Physics, Lund University,\\
S\"olvegatan 14A, SE 223-62 Lund, Sweden}
\end{center}
\vfill
\centerline{\bf Abstract}
We apply hard pion Chiral Perturbation Theory
to charmonium decays to $\pi\pi$, $KK$ and $\eta\eta$.
We first discuss why we expect to be able to provide results for the
chiral logarithms in $\chi_{c0}$ and $\chi_{c2}$ decays to two pseudoscalars
while for the decays from $J/\psi$, $\psi(nS)$ and $\chi_{c1}$ no simple
prediction is possible.
The leading chiral logarithm turns out to be absent for $\chi_{c0},\chi_{c2}\to PP$.
This result is true for all fully chiral singlet states of spin zero and two.

\vfill
\noindent{\bf PACS:}
\begin{minipage}[t]{12cm} {13.20.Gd} {Decays of $J/\psi$, $\Upsilon$ and other quarkonia}\\
{11.30.Rd} {Chiral symmetries}\\
 {12.39.Fe}{Chiral Lagrangians}
\end{minipage}
\newpage
\twocolumn
\title
{Chiral Symmetry and Charmonium Decays to Two Pseudoscalars}
\vfill
\author{Johan Bijnens \and Ilaria Jemos
\thanks{Address from 1 October 2011: University of Vienna,
Faculty of Physics, Boltzmanngasse 5, A-1090 Wien, Austria}
}
\institute{Department of Astronomy and Theoretical Physics, Lund University,\\
S\"olvegatan 14A, SE 223-62 Lund, Sweden}
\date{}
\abstract{
We apply hard pion Chiral perturbation Theory
to charmonium decays to $\pi\pi$, $KK$ and $\eta\eta$.
We first discuss why we expect to be able to provide results for the
chiral logarithms in $\chi_{c0}$ and $\chi_{c2}$ decays to two pseudoscalars
while for the decays from $J/\psi$, $\psi(nS)$ and $\chi_{c1}$ no simple
prediction is possible.
The leading chiral logarithm turns out to be absent for $\chi_{c0},\chi_{c2}\to PP$.
This result is true for all fully chiral singlet states of spin zero and two.
}
\maketitle

\section{Introduction}
\label{sect:intro}

The study of charmonium decays has seen in the last years a remarkable
progress and a renewed interest. New experimental measurements, main\-ly
coming from Belle, BES, CLEO
and E835 \cite{PDG2010} have improved existing data on several
exclusive hadronic decay channels leading to better determinations of the
charmonium branching fractions. On the theory side several heavy quarkonium
decay observables can be studied using effective field theories of QCD as
Non-Relativistic QCD (NRQCD) and its extensions \cite{Quarkonium}.
The latter is also a recent review in general of quarkonium physics.

In this letter we focus on the OZI suppressed decays of $\chi_{c0},\chi_{c2}$
into
two light pseudoscalar mesons, $\pi\pi$, $KK$, $\eta\eta$ (\chitoPP)
and we present
the arguments why we have no similar results for
 $J/\psi, \psi(nS),\chi_{c1}$ decays to the same final state.
We explore the possibility to
calculate the contributions to the decay amplitudes due to the so-called
chiral logarithms. These have the form $m^2\log{(m^2)}$, where $m$ is the mass
of a light meson\footnote{In the remainder of this letter $m$ stands
for $m_\pi$, $m_K$ and $m_\eta$.} 
and these are potentially the largest contribution
from the light quark masses. Surprisingly, we find that for \chitoPP\ these
contributions vanish. This is not necessarily the case for scalar quantities
at high $q^2$ as we show with the case of the scalar formfactor. 

Terms of the type $m^2\log(m^2)$ and other nonanalytic dependence on the
input parameters can be produced by soft meson loops \cite{LiPagels}.
The modern version of this method is Chiral Perturbation Theory (ChPT)
\cite{Weinberg0,GL1,GL2}, see also the lectures \cite{Pich,Scherer1}.
All of these applications require the octet of pseudoscalar
mesons to have soft momenta which is not the case in charmonium decays.
However, it has been argued that even for cases with pseudoscalar
mesons at large momenta there are still predictions possible.
This was first argued for $K_{\ell3}$ decays in \cite{FS}
and later argued to be more general and applied to
$K\to\pi\pi$ \cite{BC} and $B\to D,\pi,K,\eta$ vector formfactors
and the pion and kaon electromagnetic formfactors \cite{BJ1,BJ2}.
The underlying arguments were tested at two-loop level for the pion
vector and  scalar formfactor in \cite{BJ2}. We refer to this method as
hard pion ChPT (HPChPT).

The underlying argument in HPChPT is that the chiral logarithms are
coming from the soft part of all diagrams when we envisage a theory with hadrons
with an effective Lagrangian to all orders. This should be able to describe
all effects according to Weinberg's folklore theorem \cite{Weinberg0}.
The hard part of these diagrams around a particular kinematical situation
are describable by a tree level Lagrangian that is analytic in the soft physics.
Loops using the latter Lagrangian thus reproduce the nonanalytic dependence of
the underlying loop diagrams in the full theory.
The second part is then proving that we use a sufficiently complete tree level
Lagrangian. This has been done in all the previous works \cite{FS,BC,BJ1,BJ2}
by showing
that higher order operators have the same matrix element as the lowest order
operator up to terms analytic in the light pseudoscalar masses. This allowed
the determination of the chiral logarithms for the processes mentioned earlier.
The same arguments are applicable to the present decays.

In Section~\ref{sect:ChPT} we define the notation and give the lowest order
Lagrangians involving charmonium fields. Section~\ref{sect:calculation}
describes the relevant loop calculations. We note that our zero result
for the chiral logarithm is in fact valid for all fully chiral singlet
states of spin 0 and 2.
One consequence of our result is that light quark mass corrections in
the \chitoPP\ decays are not enhanced. We compare this statement
with the available experimental results in Section~\ref{sect:experiment}.

\section{Formalism}
\label{sect:ChPT}

First we shortly summarize the formalism of ChPT in
its three-flavour version \cite{GL2}. Introductions to
ChPT can be found in \cite{Pich,Scherer1}.
Hereafter we will use the same notation as in
\cite{BCE}.
In ChPT it is assumed that the spontaneous symmetry
breaking of chiral symmetry takes
place. In group theory language it has the pattern\\$SU(3)_R\times
SU(3)_L/SU(3)_R\to SU(3)_V$ . The oscillations around the vacuum are described
by the field  $u\in SU(3)$
\be\label{u}
u=\exp\left(\frac{i}{\sqrt{2}F_0}\phi\right),
\ee 
where $\phi$ is an hermitian matrix containing the pseudo Goldstone bosons,
 i.e. the light pseudoscalar
mesons
 \ba
\phi =
 \left( \begin{array}{ccc}
\frac{1}{\sqrt{2}}\pi^0  + \frac{1}{\sqrt{6}} \eta &  \pi^+ & K^+ \\
 \pi^- & -\frac{1}{\sqrt{2}}\pi^0 + \frac{1}{\sqrt{6}} \eta  & K^0 \\
 K^- & \bar{K}^0 & -\frac{2}{\sqrt{6}} \eta  \\
 \end{array} \right)\,.
\ea
The lowest order Lagrangian describing the strong interactions of the mesons
must satisfy the same symmetries of QCD and reads 
\be
\label{pilagrangian}
\mathcal{L}_{\pi \pi}^{(2)}= \frac{F^2_0}{4}  
\left( \langle u_{\mu} u^{\mu} \rangle   + \langle \chi_{+}\rangle \right),
\ee
with
\ba
 u_{\mu} &=& i\{  u^{\dag}( \partial_{\mu} - i r_{\mu}   )u
 -u ( \partial_{\mu}   -i l_{\mu}    ) u^{\dag}   \}\,,
\nonumber\\
\chi_{\pm} &=& u^{\dag} \chi u^{\dag} \pm u \chi^{\dag} u\,,
\nonumber\\
 \chi &=& 2B_0 (s+ip)\,.\nonumber
\ea
The fields $s$, $p$, $l_\mu=v_\mu-a_\mu$ and
$r_\mu=v_\mu+a_\mu$ are the standard external scalar, pseudoscalar, left- and
right-handed vector fields introduced by Gasser and Leutwyler \cite{GL1,GL2}.
$s$ includes the light quark mass matrix.

The field $u$ and $u_\mu$ transform under a chiral transformation $g_L\times g_R
\in SU(3)_L\times SU(3)_R$ as
\ba
\label{trasfrules}
u \longrightarrow g_R u h^\dagger = h u g_L^\dagger,\qquad
u_\mu\longrightarrow h u_\mu h^\dagger.
\ea
In (\ref{trasfrules}) $h$ depends on $u$, $g_L$ and $g_R$ and
is the so called compensator field.
The notation $\langle X\rangle$ stands for trace over up, down, strange.

As anticipated in Section~\ref{sect:intro} the use of ChPT is limited to
those processes where the pseudoscalar mesons are not very energetic. 
Indeed the power counting is based on an expansion in $m^2/\Lambda^2$ and in
$p^2/\Lambda^2$, where $\Lambda$ is around $1$ GeV and is the scale up to
which ChPT is believed to work. However sometimes the dependence on the light
pseudoscalar masses outside such energy regime is needed. In such cases the
extension of ChPT to hard pion ChPT is necessary. For an extended explanation
of the arguments leading to hard pion ChPT we refer the interested reader to
\cite{FS,BC,BJ1}. Here we only stress the key
features of such extension. When calculating the amplitudes of processes
involving hard pseudoscalar mesons the Feynman diagrams contain both hard
and soft lines. The
application of hard pion ChPT involves a separation of the first ones from the
latter. The hard lines get contracted into a vertex of another effective
Lagrangian. The couplings appearing there must therefore depend on the hard
quantities. The soft lines instead encode the dependence on the soft parts,
like the light masses of the pseudoscalar mesons, and are used to calculate the
chiral logarithms. Such calculation is done using a Lagrangian built up
in the same fashion as the one of standard ChPT since the hard parts must
satisfy the constraints from chiral symmetry.
However one must keep in mind that now the expansion that survives is in the
small parameter $m^2/\Lambda^2$, while
the one in momenta must be dropped. This means that adding extra derivatives
to the Lagrangian is in principle allowed since they are not suppressed by
extra powers of momentum. Fortunately it often turns out that the 
matrix elements of operators
containing these extra derivatives are
proportional to those of the lowest order operator up to
terms either analytic in $m^2$
or higher order. The coefficient of $m^2\log m^2$ is thus predictable.
It is important then that this happens in the case under
study as well. 

The arguments in \cite{FS,BC} are for a heavy to two light decay
and the exact same arguments apply to the present case of charmonium to two
light pseudoscalars.

Charmonium states are full chiral singlets, i.e. they do not transform at
all under the chiral $SU(3)_L\times SU(3)_R$. Their kinetic terms are thus
the standard kinetic terms for fields of the given spin without couplings to
pions. The results obtained here are thus valid for all full chiral
singlets.

We only study decays to two light pseudo\-scalar mesons, in principle
we could study decays to three light mesons as well but
the extension is not totally straightforward since the HPChPT should be
applied to each kinematic configuration separately.
Also for more particles the operator structure is typically much more involved.

We could as well study $J/\psi,\psi(nS)$
and $\chi_{c1}$ decays into two light pseudo\-scalar
mesons. Again the reason why we have not done this is that it would be
impossible to compare  $J/\psi\rightarrow\pi\pi$ with respect to
$J/\psi\rightarrow KK$ and similar for the others.
These two decays are caused by different
operators. The first one violates $G$-parity,\footnote{Or more general, an
$L=1$ state of two pions is in an $I=1$ state.}
it thus proceeds either through the electromagnetic interaction
or the quark mass difference $m_d-m_u$. The decays to a pair of kaons
have the same problem but instead with $U$ or $V$-spin rather than isospin.
These decays thus proceed electromagnetically or through the mass differences
$m_s-m_u$ or $m_s-m_d$.
Notice that this is also the reason
why the decay of $J/\psi$ into $\pi\pi$ is suppressed compared to the one into
$KK$. It follows that a simple comparison, as the one done in
Section~\ref{sect:calculation} for $\chi_{c0},\chi_{c2}$, is not possible.

We describe the $\chi_{c0}$ state with a chiral singlet,
scalar field $\chi_0$ and the $\chi_{c2}$ with a chiral singlet symmetric
traceless tensor field $\chi_{2\mu\nu}$ satisfying
$\chi_{2\mu\nu}=\chi_{2\nu\mu}$, $\eta^{\mu\nu}\chi_{2\mu\nu}=0$
and on-shell $p_\chi^\mu \chi_{2\mu\nu}=0$.

We
remind the reader that we are interested in calculating the dependence of the
amplitudes on $m^2$. To predict this we will stop at the chiral logarithm
level so we only calculate contributions like $m^2\log{m^2}$. Terms of order
$m^2$ are instead of higher order and thus neglected.
This means we can neglect all effects of $\chi_\pm$ except for the
contribution to the light meson masses and their effects in loop
diagrams. In particular, effects of the light quark masses
on the charmonium mass are linear in the light quark masses or higher order.

It turns out that there is only one lowest order operator for each
case
\be
\label{LOchi}
\mathcal{L}_{\chi_c} = E_1 F_0^2\,\chi_0\,\langle u^\mu u_\mu \rangle
+ E_2 F_0^2 \,\chi_2^{\mu\nu}\,\langle u_\mu u_\nu\rangle\,.
\ee
We have added a factor of $F_0^2$, the chiral limit pion decay constant,
to have the lowest order independent of $F_0$.
We have actually added a few more terms for the scalar case as an explicit check
on the HPChPT arguments.
These are
\be
\mathcal{L}_{\chi_c}^E = E_3 F_0^2\,\chi_0^2\,\langle u_\mu u^\mu\rangle
+ E_4 F_0^2\,\chi_0\, \langle \nabla^\mu u^\nu \nabla_\mu u_\nu\rangle\,. 
\ee

\section{The Calculation}
\label{sect:calculation}

\begin{figure}
\centerline{\includegraphics[width=\columnwidth]{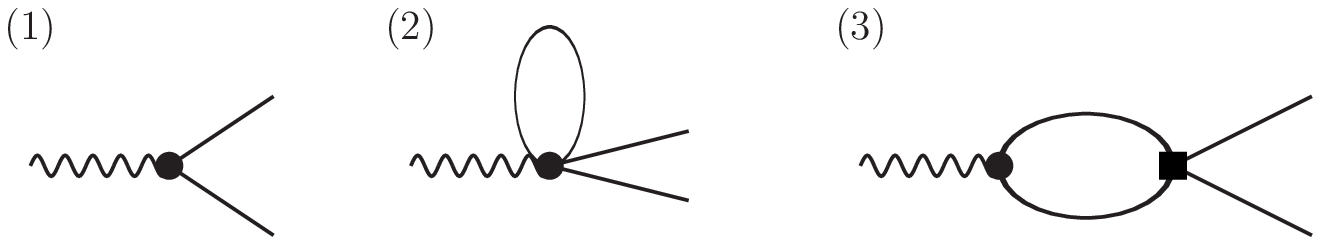}}
\caption{\label{fig:diagrams1} The tree-level (1) and the loop-diagrams
(2,3) contributing to the decays $\chitoPP$
up to the order $m^2\log m^2$.
The wiggly lines correspond to a $\chi_c$ while the continuous line are
$\pi$, $K$ or $\eta$. A round vertex correspond to an interaction from
$\mathcal{L}_{\chi_c}$ while a box is a vertex from $\mathcal{L}_\pi$.
Notice that the diagrams of type (3) do not contribute with chiral
  logarithms to the amplitudes.}
\end{figure}

The tree-level diagrams and the loop-diagrams that contribute to \chitoPP\
are shown in Figure~\ref{fig:diagrams1}. To these diagrams we also need
to add the wavefunction renormalization for the pseudoscalar meson.
The wave function renormalization for the charmonium states has no
contributions of order $m^2\log m^2$.
The tree level result from the diagram in Figure~\ref{fig:diagrams1}(1)
reads
\ba
iA(\chi_{c0}\to PP)&=& - (m_\chi^2-2m_P^2)\times
\nonumber\\&&
\left(2E_1 + E_4(m_\chi^2-2m_P^2)\right)\,,
\nonumber\\
iA(\chi_{c2}\to PP)&=& -4 E_2 p_1^\mu p_2^\nu T_{\chi\mu\nu}\,.
\ea
The amplitude is the same for the final states
$\pi^+\pi^-,K^+K^-,K^0\overline{K^0}$ and $\eta\eta$ when the meson
mass $m_P$ is chosen to be the appropriate one.
$T_{\chi\mu\nu}$ is the polarization vector of $\chi_{c2}$ and $p_1$ and
$p_2$ are the fourmomenta of the two mesons in the final state.
Here we also see that the example of a higher derivative term
containing $E_4$ indeed gives a result proportional
to the lowest term up to corrections of order $m^2$.

When we go over the powercounting arguments, only soft pseudoscalar
propagators where no derivatives act on the pseudoscalar fields associated
with that propagator can give contributions of order $m^2\log m^2$.
As a consequence only pseudoscalar meson wave function renormalization and
the diagram in Figure~\ref{fig:diagrams1}(2) can contribute terms of order
$m^2\log m^2$ and we find that these two contributions exactly cancel for all
decays considered. We indeed find that the diagram in
Figure~\ref{fig:diagrams1}(3) does not contribute terms of order $m^2\log m^2$.

\begin{figure}
\centerline{\includegraphics[width=0.4\columnwidth]{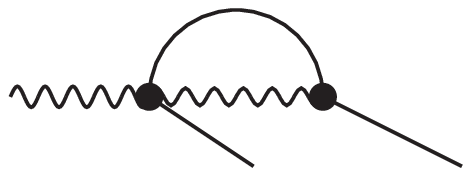}}
\caption{
\label{fig:diagrams2} 
An example diagram involving a propagator of the heavy particle in the loop.
These do not contribute to the chiral logarithms of
order $m^2\log m^2$.}
\end{figure}

In baryon ChPT, and other extensions with heavy particles,
one can also get logarithmic contributions from diagrams
involving propagators of the heavy particle. The powercounting argument
together with the fact that all interactions terms have derivatives show
that these should not be present here. We have explicitly confirmed this with
the diagram of Figure~\ref{fig:diagrams2}.
Putting in the vertices with $E_1$ and $E_3$ we indeed find no
contributions of order $m^2\log m^2$.

We conclude this part with a final remark. The above discussion is valid
assuming that other charmonium states are far enough in mass, so that such
states cannot appear as virtual particles in the loops, i.e. their contributions
are sufficiently hard that they can be described by the same tree level
Lagrangian (\ref{LOchi}).

We were somewhat surprised to find that there were no chiral logarithms of
leading order in these decays, and more, since the only assumption that goes
in is the HPChPT and the fact that we have a chiral singlet field.
The same arguments go through for the energy momentum
tensor $T^{\mu\nu}$ which is a spin two chiral singlet.
The expressions for the matrix element of the energy momentum tensor between
pseudoscalar states $\langle P| T^{\mu\nu} | P\rangle$ are known to one-loop
order in ChPT \cite{DL}. Our result should predict the $m^2\log m^2$ parts
of their result in the limit $q^2\gg m^2$. Expanding (29) and (33) in
\cite{DL} in $m^2/q^2$ we indeed find that there are no logarithms
of order $m^2\log m^2$ appearing there.

In earlier work we have found many instances of nonzero chiral logarithms in
HPChPT. Most of these were for the vector formfactor but we did
find a nonzero result for the pion scalar formfactor
$\langle\pi|\bar u u+\bar d d|\pi\rangle$ at large $q^2$ in two-flavour HPChPT.
Note that $\bar u u+\bar d d$ is a singlet under $SU(2)_V$ but not
under the full chiral $SU(2)_L\times SU(2)_R$ so the calculations given
above do not restrict the chiral logarithms at large $q^2$ here.

\section[Comparison with experiment]
{Comparison with experiment for \chitoPP}
\label{sect:experiment}

The fact that the chiral logarithms of order\\$m^2\log m^2$ vanish means
that the leading term in the expansion of $m^2$ of the amplitudes vanish.
One could thus expect that the $SU(3)_V$ breaking in \chitoPP\ should be
somewhat smaller than the ``usual'' 20\% 
as e.g. in $F_K/F_\pi$ or $m_\Lambda/m_p$.
Terms of order $m^2$ are however not predicted, so 
a very clear statement that $SU(3)_V$ breaking effects are small
is not possible.

Let us however see how well the measured amplitudes live up to the ``small
$SU(3)_V$ breaking effects.''
We take the input data from \cite{PDG2010}. These are given in
Table~\ref{tab:data}.
\begin{table*}
\begin{center}
\begin{tabular}{|c|c|c|c|c|}
\hline
 &\multicolumn{2}{c|}{$\chi_{c0}$} & 
  \multicolumn{2}{c|}{$\chi_{c2}$}\\
\hline
Mass &\multicolumn{2}{c|}{$3414.75\pm0.31$~MeV}
     &\multicolumn{2}{c|}{$3556.20\pm0.09$~MeV}\\
Width&\multicolumn{2}{c|}{$10.4\pm0.6$~MeV}
     &\multicolumn{2}{c|}{$1.97\pm0.11$~MeV}\\
\hline
Final state & $10^3\,$BR & $10^{10}\,G_0[\mathrm{MeV}^{-5/2}]$ 
            &$10^3\,$BR & $10^{10}G_2[\mathrm{MeV}^{-5/2}]$ \\
\hline
$\pi\pi$    & $8.5 \pm0.4   $   & $3.15 \pm0.07 $  
            & $ 2.42\pm0.13 $ & $3.04\pm0.08 $\\
$K^+ K^-$     & $6.06 \pm0.35 $ & $3.45 \pm0.10 $
              & $1.09 \pm0.08 $ & $2.74 \pm0.10 $ \\
$K^0_S K^0_S$  & $3.15 \pm0.18 $ & $3.52 \pm0.10 $ 
              & $0.58 \pm0.05 $   & $2.83 \pm0.12 $  \\
$\eta\eta$    & $3.03 \pm0.21$ & $2.48 \pm0.09 $ 
              & $0.59 \pm0.05 $   & $2.06 \pm 0.09$ \\
$\eta^\prime\eta^\prime$ & $2.02 \pm0.22$ & $2.43 \pm0.13 $ 
              & $<0.11$   & $<1.2$ \\
\hline
\end{tabular}
\end{center}
\caption{\label{tab:data} The Experimental results for the decays \chitoPP\ and
the resulting factors corrected for the known $m^2$ effects.}
\end{table*}
It should be taken into account that the $\pi\pi$ is the sum
of $\pi^+\pi^-$ and $\pi^0\pi^0$ final state. Isospin predicts that the
$\pi^0\pi^0$ is half the  $\pi^+\pi^-$ state and the data that go into the
PDG average are compatible with this. We thus need to multiply it by
$2/3$ to get the $\pi^+\pi^-$ result.
Similarly, the $K^0_SK^0_S$ final
state is half of the $K^0\overline{K^0}$ final state. Here we need
to multiply by two to obtain the full final state.

One question is how we deal with factors of $m_P^2$ that we know are present.
We include the phase space correction and note that for $\chi_{c0}$
the amplitude always contains $p_1\cdot p_2=(m_\chi^2-2m_P^2)/2$.
The phase space contains the factor $|\vec p_1|=\sqrt{m_\chi^2-4m_P^2}/2$.
For the decays of the scalar we thus define
\be
G_0 = \sqrt{BR/|\vec p_1|}/(p_1.p_2)\,.
\ee
The same question arises for the $\chi_{c2}$. Here the amplitude
must always contain $\chi_2^{\mu\nu}p_{1\mu}p_{2\nu}$.
The field $\chi_2^{\mu\nu}$ must be replaced by its polarization tensor
$T_\chi^{\mu\nu}$ and the amplitude squared and
averaged over.
The formula to perform such a sum is
\ba\label{sumpol}
\sum_{\rm polarizations}T^{\mu\nu}T^{*\alpha\beta}&=&\frac{1}{2}
\Bigg(K^{\mu\alpha}K^{\nu\beta}+
K^{\mu\beta}K^{\nu\alpha}
\nonumber\\&&
-\frac{2}{3}K^{\mu\nu}K^{\alpha\beta}\Bigg)
\ea
where $K^{\mu\nu}=-g^{\mu\nu}+q^\mu q^\nu /m^2_{\chi}$ and $q^\mu$ is
the four-momentum of the $\chi_{c2}$ particle.
The contribution thus always contains a factor
\be
\frac{1}{5}\sum_{pol}T_\chi^{\mu\nu}p_{1\mu}p_{2\nu}
T_\chi^{*\alpha\beta}p_{1\alpha}p_{2\beta}
= \frac{1}{30}\left(m_\chi^2-4m_P^2\right)^2\,.
\ee
Alternatively we can choose the final state configuration with
$p_1=(E_P,0,0,|\vec p_1|)$ and $p_2=(E_P,0,0,-|\vec p_1|)$ and choose
an explicit set of five orthogonal polarization tensors satisfying
$p_\chi^\mu T_{\chi\mu\nu}=0,$ $T_{\chi\mu}^\mu=0$, 
$T_{\chi\mu\nu}=T_{\chi\nu\mu}$ and\\
$T_{\chi\mu\nu}^{(a)}T^{*(b)\mu\nu}_\chi = \delta^{ab}$ which shows
that the amplitude squared
always contains a $|\vec p_1|^4$. We thus define a normalized factor
also for the $\chi_{c2}$ decays via
\be
G_2 = \sqrt{BR/|\vec p_1|}/|\vec p_1|^2\,.
\ee

Looking at the columns $G_0$ and $G_2$ for the decays to pions and kaons we see
indeed that $SU(3)_V$ breaking is somewhat smaller than usual,
about 10\% for both $\chi_{c0}$ and $\chi_{c2}$ decays.
For the decays to $\eta\eta$ we get a decent agreement in both cases
but the\footnote{The arguments given in this paper are only applicable to the
$\eta^\prime$ in the large $N_c$ limit where this state also becomes
a pseudo Goldstone boson.}
$\eta^\prime\eta^\prime$ is not so good for the $\chi_{c2}$ decay.

\section{Conclusions}
\label{sect:conclusions}

In this letter we have calculated the chiral logaritms for the decays
$\chi_{c0},\chi_{c2}$ to two light pseudoscalars. We have found that they vanish
and that this is a general result for all such decays of heavy chiral singlet
states of spin 0 and 2. We checked our result against the known result
for the energy momentum tensor in ChPT and showed using the example of the
scalar formfactors that the chiral logarithms do not vanish at high $q^2$
in all cases.

Our result implies that the $SU(3)_V$ corrections are expected to be ``small''
in these decays and we have compared our results with the available
experimental data and find that the corrections are reasonably small.

\section*{Acknowledgments}

This work is supported in part by the European Community-Research
Infrastructure Integrating Activity ``Study of Strongly Interacting Matter'' 
(HadronPhysics2, Grant Agreement n. 227431)
and the Swedish Research Council grants 621-2008-4074 and 621-2010-3326.
This work used FORM \cite{Vermaseren:2000nd}.

\end{document}